\documentclass[
onecolumn,
superscriptaddress,
 amsmath,
 amssymb,
 aip,
apl,
]{revtex4-2}

\usepackage[english]{babel}
\usepackage[utf8]{inputenc}
\usepackage{xr-hyper}
\usepackage{float}
\usepackage{dcolumn}
\usepackage{bm}
\usepackage[colorlinks]{hyperref}
\usepackage[T1]{fontenc}
\usepackage[dvips]{graphicx}
\usepackage[dvipsnames]{xcolor}
\usepackage[font=small,labelfont=bf,justification=justified,format=plain,singlelinecheck=false]{caption}
\usepackage{url}
\usepackage{ulem}
\usepackage{braket}
\usepackage{array}

\usepackage{tikzducks}

\makeatletter
\newcommand*{\addFileDependency}[1]{
  \typeout{(#1)}
  \@addtofilelist{#1}
  \IfFileExists{#1}{}{\typeout{No file #1.}}
}
\makeatother

\usepackage{subcaption}

\hypersetup{filecolor=blue}
\hypersetup{citecolor=blue}
\hypersetup{urlcolor=blue}
\hypersetup{linkcolor=blue}

\usepackage{hyperref}
\usepackage[mathlines]{lineno}

\begin{document}

\title{\textcolor{blue}{Inverse Rashba-Edelstein THz emission modulation induced by ferroelectricity in CoFeB/PtSe$_2$/MoSe$_2$//LiNbO$_3$ systems}}

\author{S.~Massabeau}
\affiliation{Laboratoire Albert Fert, CNRS, Thales, Université Paris-Saclay, 91767 Palaiseau, France}

\author{O.~Paull}
\affiliation{Laboratoire Albert Fert, CNRS, Thales, Université Paris-Saclay, 91767 Palaiseau, France}

\author{A.~Pezo}
\affiliation{Laboratoire Albert Fert, CNRS, Thales, Université Paris-Saclay, 91767 Palaiseau, France}

\author{F.~Miljevic}
\affiliation{Laboratoire Albert Fert, CNRS, Thales, Université Paris-Saclay, 91767 Palaiseau, France}

\author{M.~Mičica}
\affiliation{Laboratoire de Physique de l’Ecole Normale Supérieure, ENS, Université PSL, CNRS, Sorbonne Université,
Université Paris Cité, F-75005 Paris, France}

\author{A. Grisard}
\affiliation{Thales Research \& Technology, 91767 Palaiseau, France}

\author{P. Morfin}
\affiliation{Laboratoire de Physique de l’Ecole Normale Supérieure, ENS, Université PSL, CNRS, Sorbonne Université,
Université Paris Cité, F-75005 Paris, France}

\author{R. Lebrun}
\affiliation{Thales Research \& Technology, 91767 Palaiseau, France}

\author{H.~Jaffrès}
\affiliation{Laboratoire Albert Fert, CNRS, Thales, Université Paris-Saclay, 91767 Palaiseau, France}

\author{S. Dhillon}
\affiliation{Laboratoire de Physique de l’Ecole Normale Supérieure, ENS, Université PSL, CNRS, Sorbonne Université,
Université Paris Cité, F-75005 Paris, France}

\author{J.-M.~George}
\affiliation{Laboratoire Albert Fert, CNRS, Thales, Université Paris-Saclay, 91767 Palaiseau, France}

\author{M.~Jamet}
\affiliation{Univ.~Grenoble Alpes, CEA, CNRS, Grenoble INP, IRIG-SPINTEC, F-38000 Grenoble, France}

\author{M.~Bibes}
\affiliation{Laboratoire Albert Fert, CNRS, Thales, Université Paris-Saclay, 91767 Palaiseau, France}

\date{\today}

\newpage

\begin{abstract}
Spintronic Terahertz emitters, based on optically triggered spin-to-charge interconversion processes, have recently emerged as novel route towards compact and efficient THz sources. Yet, the next challenge for further technologically-relevant devices remains to modulate the emission, with low-energy consumption operation. To this aim, ferroelectric materials coupled to active spin-orbit layers such as two-dimensional transition metal dichalcogenides are suitable candidates. In this work, we present the realization of a large area heterostructure of CoFeB/PtSe$_2$/MoSe$_2$ on a bi-domain LiNbO$_3$ substrate. Using THz time-domain spectroscopy, we show that the ferroelectric polarization direction induces a sizeable modulation of the THz emission. We rationalise these experimental results by using band structure and spin accumulation calculations that are consistent with an interfacial spin-to-charge conversion mediated by inverse Rashba-Edelstein effect at the MoSe$_2$/PtSe$_2$ interface and being tuned by ferroelectricity in the adjacent LiNbO$_3$ surface. This work points out the relevance of field effect spin-orbit architectures for novel THz technologies.

\end{abstract}

\maketitle

\newpage
\section*{\label{sec:Intro} Introduction}
Over the last decade, huge efforts have been put forward in the development of novel kind of THz sources based on ultrafast spin-to-charge interconversion, thus referred as spintronic THz emitters. These emitters are defined, in their simplest form, by a heterostructure made of nanometer thick ferromagnetic (FM)/non-magnetic (NM) junctions. When excited by an ultrashort optical pulse, an out-of-equilibrium spin current is generated in the FM layer that is diffused and converted to a transient charge current on the NM side. This layer acts as an active spin-orbit layer and subsequently leads to the generation of a linearly polarized THz pulse~\cite{seifert_spintronic_2022}. The spin-to-charge conversion (SCC) is driven by the inverse spin-Hall effect in the case of bulk spin-orbit materials like heavy metals~\cite{seifert_efficient_2016}, and/or mediated by the inverse Rashba-Edelstein effect (IREE) at the FM/NM interface in the presence of a spin-polarized texture~\cite{jungfleisch_control_2018,zhou_broadband_2018}, e.g. on the surface of topological insulators~\cite{chen_currentenhanced_2019,rongione_spinmomentum_2023,rho_exceptional_2023}.

Such ultrathin emitters exhibit broadband and gap-less THz emission spectrum over a few tens of THz, and their polarization direction can be easily controlled by a small applied magnetic field oriented in the sample plane. Many works have been subsequently carried out in order to increase the absolute emitter power by optimizing multisequenced stacks, thicknesses and the nature of FM and NM in the heterostructure~\cite{gueckstock_terahertz_2021,hawecker_spintronic_2022}. This progress has led to optimized emitters with performance already surpassing usual THz sources such as non-linear crystals~\cite{rouzegar_broadband_2023}, making spintronic emitters concrete candidates for the development of THz technologies.
In particular, the THz spectral band is expected to be of paramount importance in the development of the next generation telecommunication standard~\cite{leitenstorfer_2023_2023}, with various applications such as high speed wireless communication, nano-networks and networks-on-a-chip. However, these applications require controlling the emission with an external degree of freedom in order to encode information. Due to the recent emergence of spintronic emitters, only a few works have addressed this crucial need towards different strategies like patterning the emitter itself or using complex optical excitation schemes~\cite{ji_tunable_2023,nkeck_parallel_2022}. Nevertheless, such methods remain limited as they permanently affect the design of the emitter or require bulky and energy-consuming equipments, thus reducing device scalability and adaptability.

Another relevant approach relies on the combination of ferroelectric materials with 2D electron gas, such as one made of an ultrathin 2D material with strong spin-orbit coupling~\cite{ning_challenges_2022}. These field effect spin-orbit (FESO) devices allow to control the spin-orbit coupling in the 2D layer via ferroelectricity, which could be switched externally by the application of a gate voltage~\cite{noel_non-volatile_2020}. It is thus particularly appealing for the development of low-consumption THz devices to couple such non-volatile voltage-controlled switching with a source of spin-current, in order to generate and modulate THz emission.
The key ingredients for this aim are therefore (i) a stabilized and technologically relevant ferroelectric material (ii) a suitable 2D material with spin textures modulated by electric field, which could be easily grown or transferred onto the ferroelectric material and (iii) an entire heterostructure with a mm-sized active surface for an easy activation by optical pulses. \\

On one hand, lithium niobate (LiNbO$_3$) is a well known ferroelectric material, hosting a polarization pointing along the c-axis of the hexagonal unit cell. Owing to its THz emission properties~\cite{hirori_single-cycle_2011,zhang_14mj_2021}, it has been extensively studied at THz frequencies, and is widely used in optoelectronic devices for telecommunication, allowing high modulation speed up to several tens of GHz in the C-band~\cite{rao_high-performance_2016}. On the other hand, ultrathin layers of transition metal dichalcogenide (TMD) such as PtSe$_2$ possess several advantages to tackle the aforementioned (ii) requirement. The electronic and optical properties of such van der Walls (vdW) heterostructures can be tuned by the number of stacked layers~\cite{abdukayumov_atomiclayer_2024}, they possess strong spin-orbit coupling and their band structure can be strongly modified in contact with a ferroelectric material~\cite{salazar_visualizing_2022}. Additionally, single crystal PtSe$_2$ of controlled thickness can be grown by molecular beam epitaxy (MBE) over waver-scale dimensions, in contrast with exfoliated layers. Yet, a direct growth of high quality TMD layers over large ferroelectric substrates remains challenging in terms of chemical compatibilities, and secondly as growth temperature conditions may alter ferroelectricity. \\

In this work, we present the realization of a field effect spin-orbit structure made of CoFeB/PtSe$_2$/MoSe$_2$ on a bi-domain LiNbO$_3$ substrate, allowing to modulate THz emission via ferroelectricity while excited by ultrashort optical pulses. The samples were prepared by transferring high quality MBE-grown TMD layers on a single LiNbO$_3$ wafer with cm-scale out-of-plane pre-poled areas in opposite directions. Using THz time-domain spectroscopy, we first show that the poled areas can be clearly identified in the non-magnetic THz signature, and that the ferroelectric polarization direction leads to a sizeable modulation of the magnetic THz emission. We then interpret this result by a spin-to-charge conversion mediated by IREE, with the existence of ferroelectrically-controlled Rashba states at the PtSe$_2$/MoSe$_2$ interface. Using Density Functional Theory calculations, we finally demonstrate that the ferroelectricity direction shifts accordingly the TMD bands, thus supporting our experimental results and subsequent analysis.

\section{\label{sec:Fab} Fabrication and characterization.}

\begin{figure}[ht]
  \centering
  \includegraphics[width=16cm]{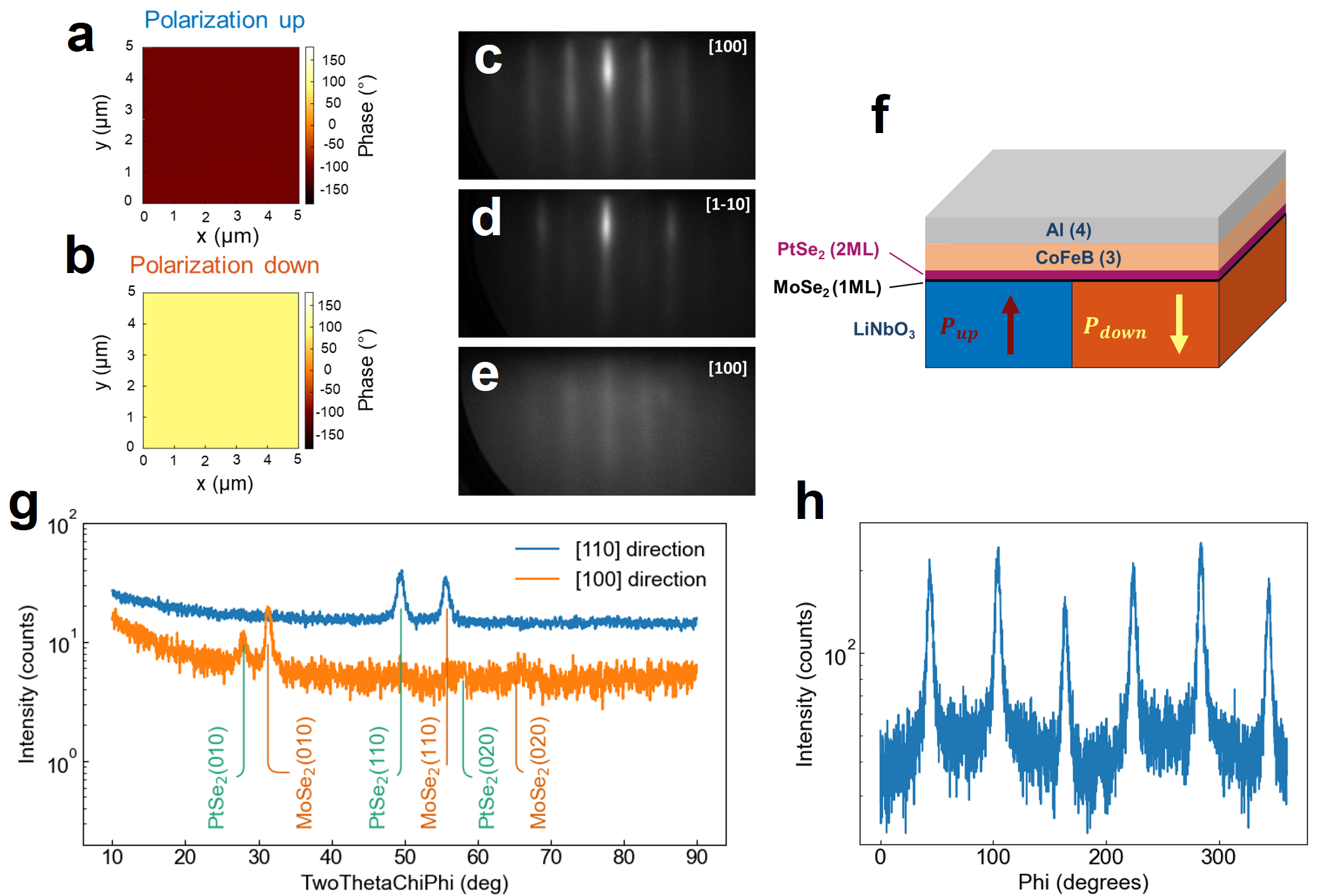}
  \caption{(a-b) Piezoresponse force microscopy data of the substrate showing the polarisation up (a) and polarisation down (b). (c-d) Reflected high-energy electron diffraction (RHEED) images along the (c) [100] and (d) [1-10] crystallographic directions directly after growth. (e) RHEED image along the [100] direction after the transfer process of the 2D layer from the Mica to the ferroelectric LiNbO$_3$ substrate. (f) Schematic of the full heterostructure after 2D material transfer and ferromagnetic layer deposition. Numbers in parentheses express the number of monolayers (thickness in nm) for 2D (bulk) materials, respectively. (g) Radial in-plane x-ray diffraction measurements along the [100] (orange) and [110] (blue) directions of the 2D layer after transferring onto the LiNbO$_3$ substrate, highlighting that crystalline structure is identical to the one on Mica after epitaxial growth. (h) Azimuthal scans of the PtSe$_2$ [110] diffraction peak, showing clear 6-fold symmetry consistent with high quality single crystal growth that is maintained after the transfer process.}
  \label{Fig1}
\end{figure}

Samples were first prepared from a LiNbO$_3$ wafer provided by Covesion, with pre-poled out-of-plane polarization areas along opposite directions. The wafer was cut into 10x10 mm² coupons, with half the area hosting a polarization towards the surface (up area) and half towards its back (down area).  As depicted in Fig.~\ref{Fig1}a-b, piezoresponse force microscopy characterization confirms the existence of well-defined up and down polarization areas over a large scale, with a very low roughness surface (RMS=0.33 nm, see Suppl. Info.~\ref{sec:S_surface}). \\

The 2D layers constituting the MoSe$_2$/PtSe$_2$ vdW heterostructure were grown using molecular beam epitaxy. Molybdenum and Platinum were evaporated with a rastered electron beam and directed onto a Mica substrate under Selenium flux. A calibrated quartz crystal balance was used to accurately measure the layer thickness such that 1 monolayer (1ML) of MoSe$_2$ could be grown, followed by 2 monolayers (2ML) of PtSe$_2$, and the quality of the growth was monitored in-situ with reflected high-energy electron diffraction (RHEED). Fig.~\ref{Fig1}c shows the RHEED pattern along the [100] direction directly after growth, with clear diffraction rods indicating true 2-dimensional growth of the TMD layers. The complementary diffraction pattern in Fig.~\ref{Fig1}d indicates anisotropy in the diffracted signal and is testament to the single crystal nature of the 2-dimensional growth by MBE.
The process to delaminate the 2D material in an aqueous environment that allows for cm-scale layers to be transferred from one substrate to another, has been outlined elsewhere~\cite{dau_van_2019,salazar_visualizing_2022}. This procedure was carried out to transfer the PtSe$_2$(2 ML)/MoSe$_2$(1 ML) heterostructure onto the LiNbO$_3$ substrate, thus ensuring the same fabrication conditions for up and down areas. The sample is followingly annealed by slowly ramping up to around 250°C in increments and held there for around an hour before cooling down, allowing to improve the surface flatness between LiNbO$_3$ and the 2D layers while preventing alteration of the ferroelectric substrate. Finally, a 3 nm-thick ferromagnetic layer made of CoFeB is deposited by sputtering without breaking vacuum, followed by 4 nm of Al to prevent oxidation, defining the heterostructure depicted in Fig.~\ref{Fig1}f. We confirm that the full stack is preserved after the transfer by in-plane X-ray diffraction measurements, which show the main characteristic diffraction peaks for each layers (Fig.~\ref{Fig1}g). An azimuthal scan of the PtSe$_2$ [110] diffraction peak (Fig.~\ref{Fig1}h), showing a clear six-fold symmetry, finally confirms that crystallinity is preserved after the transfer process.
\section{\label{sec:THz_nonmag} THz-time domain spectroscopy characterization.}
We now turn to the optical characterization of our sample, using standard THz time-domain spectroscopy technique in transmission mode. The heterostructure is excited from the substrate side (LiNbO$_3$) at normal incidence by linearly polarized 80 fs-long pulses centered at 800 nm, delivered by a Yb-doped solid state femtosecond laser coupled to an optical parametric amplifier, while applying a small in-plane magnetic field ($\sim$20mT). The repetition rate is 150 kHz and the beam waist on the sample is 250 µm. The emitted THz pulses are then detected orthogonally to the magnetic field by electro-optic sampling~\cite{dang_ultrafast_2020} with a 200 µm-thick ZnTe crystal, using the main output of the oscillator (130 fs pulses centered at 1030 nm).

Note that there are two possible sources of THz radiation within the heterostructure, each relying on different mechanisms. Under near infrared ultrafast excitation, the ferroelectric layer leads to THz emission via optical rectification, whereas the ultrafast demagnetization of the CoFeB is expected to generate THz via SCC. Pumping the sample through its substrate and detecting the emission in transmission geometry thus avoids the THz signature originating from SCC to be altered by LiNbO$_3$ absorption at THz frequencies~\cite{li_terahertz_2009}.

In order to extract the underlying symmetries of the system, we probe up and down ferroelectric areas while rotating the sample along its azimuthal axis, for two opposite directions of the magnetic field (i.e. setting the magnetization of CoFeB layer in the direction of the field). Such measurements allow to extract the non-magnetic and magnetic contributions to the THz emission by performing the sum and difference of the time traces, respectively~\cite{rongione_ultrafast_2022}.

\begin{figure}[ht]
  \centering
  \includegraphics[width=16cm]{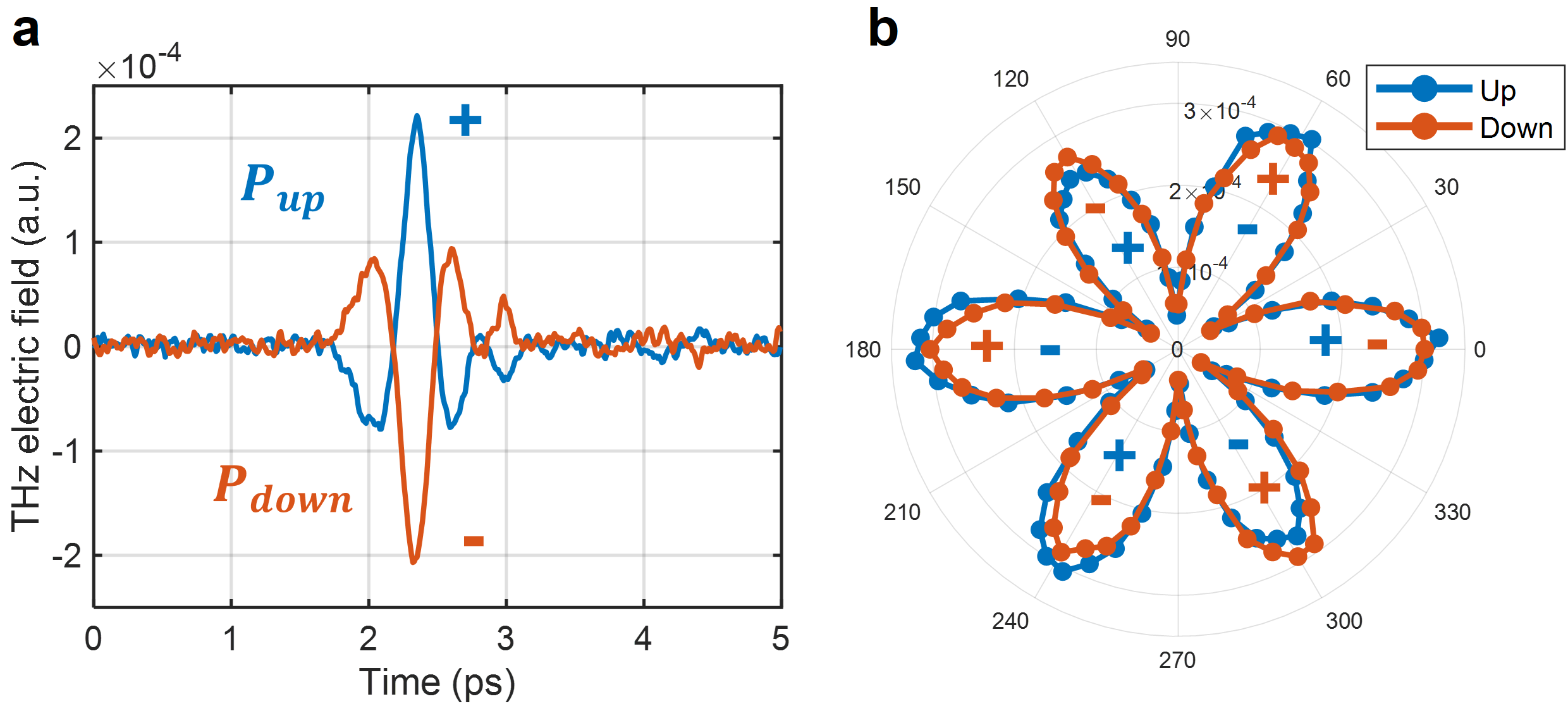}
  \caption{(a) Time traces of non-magnetic THz contribution, taken at maximum emission angle, for up (blue) and down (red) polarization areas of the LiNbO$_3$ substrate, showing a full phase reversal. We define a positive (+)/negative (-) phase when the absolute maximum is positive/negative, respectively. (b) Peak-to-peak amplitude of the non-magnetic contribution to THz emission, reported on a polar plot as a function of the sample angle, for both polarization states. The (+/-) sign on each lobe refers to the phase (positive/negative) of the corresponding time trace.}
  \label{Fig2}
\end{figure}

We first focus on the non-magnetic contribution to the THz emission. In Fig.~\ref{Fig2}a, the measured THz time trace is exhibiting a full phase reversal by switching between up and down ferroelectric areas. As the non-magnetic part originates from a non-linear second order effect ($\chi^{(2)}$ of LiNbO$_3$)~\cite{carletti_nonlinear_2023}, reversing the ferroelectric polarization is equivalent to flip the sample, i.e. inverting the dielectric tensor, thus reversing the phase of THz emission. This property is also preserved for any azimuthal angle, as depicted in Fig.~\ref{Fig2}b. While rotating the sample and reporting the peak-to-peak THz amplitude on a polar plot, we recover a three-fold symmetry for both pre-poled areas, characteristic of optical rectification in Z-cut LiNbO$_3$ crystals, with reversed phase in each lobe. Thanks to this observation, we can clearly identify the ferroelectric state of the probed area. \\

\begin{figure}[ht]
  \centering
  \includegraphics[width=16.2cm]{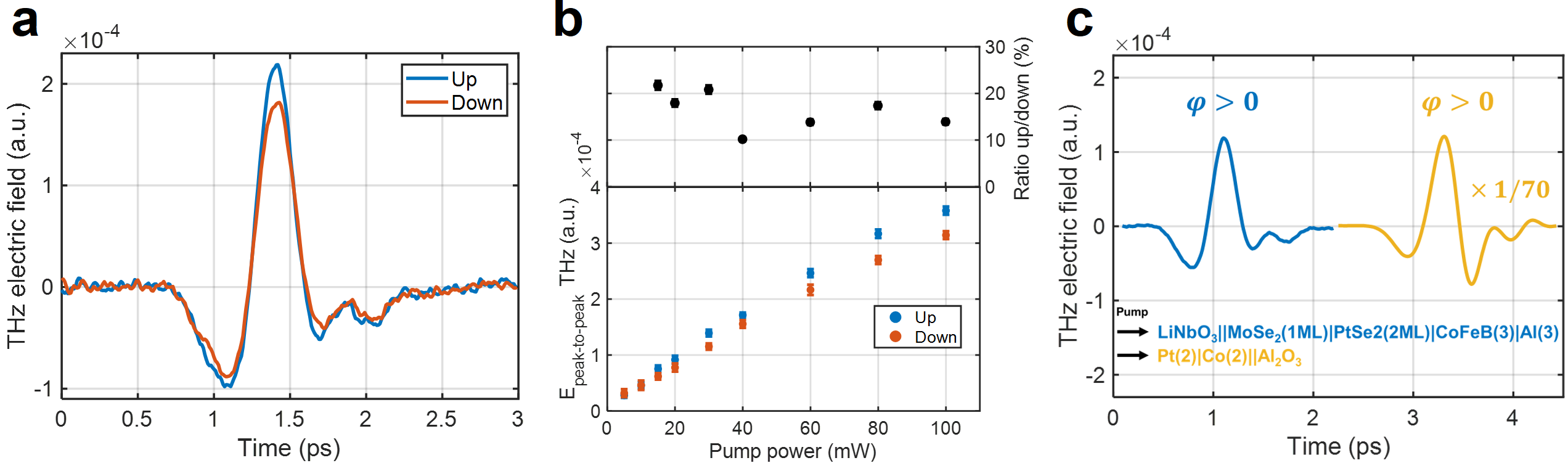}
  \caption{(a) Temporal traces of magnetic THz contribution, taken at 80 mW for up and down areas. (b) Down panel: Peak-to-peak amplitude of the magnetic contribution for up and down areas, as a function of the incident infrared pump power. Error bars are calculated using time-resolved signal standard deviation over a 500 fs time window before pulse detection. Upper panel: Calculated ratio between the peak-to-peak amplitudes in up polarized and down polarized areas in $\%$ (ratio up/down) as a function of incident pump power. (c) Temporal traces of magnetic THz contribution obtained from the heterostructure in up configuration (blue) and a standard spintronic emitter in the same experimental conditions and magnetic field. Both samples show the same conversion sign, and an emission 70 times larger for the standard emitter. Curves are shifted in time for clarity.}
  \label{Fig3}
\end{figure}

We now move to the magnetic part of the THz emission. The measured temporal THz profiles obtained from the excitation of up and down ferroelectric areas are displayed in Fig.~\ref{Fig3}a. The two time traces have the same positive phase, and we observe a sizable increase of the THz emission in the up-poled area compared to the down one. This result shows that the ultrafast spin current generated in the FM layer under illumination is efficiently converted into a charge current into adjacent layers, in the vicinity of LiNbO$_3$/2D interface. Note that in that case, the SCC process is isotropic as expected~\cite{rongione_spinmomentum_2023,rongione_ultrafast_2022,abdukayumov_atomiclayer_2024}. The influence of incident power is then investigated to further characterize the magnetic emission process, as depicted in Fig.~\ref{Fig3}b. The power dependence shows a typical behavior for SCC, linear at low power and starting to saturate above 80 mW. Interestingly, the magnetic THz contribution remains always higher for the up area over the entire power range, corroborating the influence of ferroelectricity direction on SCC efficiency. We evaluate the induced ferroelectric modulation on SCC to be about 15-20$\%$ (see top of Fig.~\ref{Fig3}b).

In order to gain more insight into the source of SCC, the THz emission amplitude and phase of our stack are further compared to a standard spintronic THz emitter made of Pt/Co grown on a sapphire substrate (Fig.~\ref{Fig3}c). Since the conversion occurs in the proximity of LiNbO$_3$/2D interface - which is further supported by our calculations shown afterwards - both samples are probed while keeping the same stacking order between spin convertor (Pt for standard emitter) and spin injector (FM layer), with respect to incident optical excitation. We observe a positive phase for the two systems, and the amplitude is about 70 times larger for Pt/Co compared to our heterostructure. 
According to previous experimental reports on CoFeB/PtSe$_2$ as a function of the TMD thickness~\cite{abdukayumov_atomiclayer_2024}, the THz magnetic emission is reduced by a factor 10 by reducing the number of TMD MLs from 10 to 2, transitioning from a SCC mediated by ISHE to IREE. Additionally, Pt and PtSe$_2$ share the same conversion sign with a reduced emission by a factor 7 considering the same Pt thickness. Taking into account the sign and amplitude of THz emission, our results thus strongly indicate that the conversion mainly occurs in the vicinity of PtSe$_2$ through IREE interfacial conversion.

\section{\label{sec:DFT} Density Functionnal Theory calculations.}

\begin{figure}[ht]
  \centering
  \includegraphics[width=16.2cm]{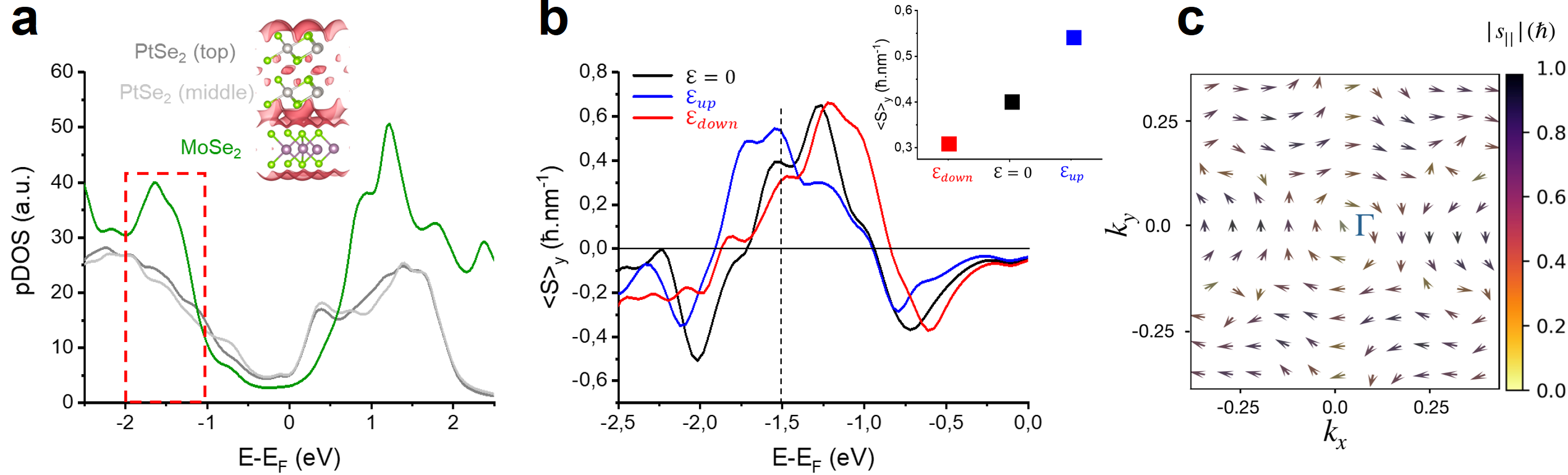}
  \caption{(a) Projected density of states (pDOS) on PtSe$_2$ top (dark grey), middle layer (light grey) and MoSe$_2$ (green). Inset: Charge density for the trilayer at the isosurface 0.01 e/{\AA}$^{3}$, displaying the strong hybridization between MoSe$_2$ and PtSe$_2$. The interlayer distances are set according to~\cite{Xu2024,Kandemir2018}. (b) Spin accumulation as a function of the energy for the trilayer, at equilibrium (black) and while applying an external electric field $\mathcal{E}$=0.25 V/{\AA} pointing up (blue) and down (red). Inset: Spin accumulation at an energy cut of E=E$_F$-1.5 eV (vertical dashed line), showing a higher value for the up case compared to the down one. (c) Spin texture at an energy cut of E=E$_F$-1.5 eV (vertical dashed line in (b)) showing Rashba states.}
  \label{Fig4}
\end{figure}

We now turn to Density Functional Theory of SCC phenomena in these systems. The band structures, spin texture and SCC linear response (Rashba-Edelstein tensor) were calculated by a home-made procedure using the Hamiltonian and overlap matrices, based on methods and equations described in Suplementary Information~\ref{sec:S_spin_acc}. Our first major results suggest that the SCC mainly occurs at the interface between MoSe$_2$ and PtSe$_2$, from states lying in a finite energy window of the valence band, as discussed in the following. In Fig.~\ref{Fig4}a, we first refer to the projected density of states on each layers forming the vdW heterostructure, extracted from the projected band structure (see Supl. Info.~\ref{sec:S_band_structure}). We recover the main feature of each layer, namely the presence of semi-conducting gap, together with an increase of MoSe2 density of states in an energy window between -2 and -1 eV (dashed red rectangle). The charge density map in the inset of Fig.~\ref{Fig4}a additionally highlights the conversion location in the heterostructure, showing a strong hybridization between MoSe$_2$ and PtSe$_2$ layers. Since MoSe$_2$ takes part to the inversion symmetry breaking, it lifts the degeneracy present in PtSe$_2$ leading to a Rashba effect as already reported in the case of PtSe$_2$/Gr bilayers~\cite{abdukayumov_atomiclayer_2024}, in which SCC also occurs in the valence band.

In order to quantify the Rashba effect modulation induced by ferroelectricity, we evaluate in Fig.~\ref{Fig4}b the spin accumulation response, related to the Rashba Edelstein tensor and thus THz emission, at equilibrium and for an electric field pointing either up or down, in agreement with our experiment. The electric field amplitude $\mathcal{E}$ is set to 0.25 V/{\AA}, taking into account that LiNbO$_3$ has a larger ferroelectric polarization compared to BaTiO$_{3}$ and CaO~\cite{vyas_first-principles_2023,novak_effect_2014}. In the absence of any additional polarization field ($\mathcal{E}$=0), a strong and positive spin accumulation is observed in the previously identified energy window, i.e. in the valence band. In the presence of an electric field pointing up (respectively down), the spin accumulation remains positive and is shifted towards lower energy (respectively higher energy), resulting in a total energy shift of about 0.32 eV. Such shift is interpreted from the semi-conducting nature of the layers, which prevents field screening from the substrate, and is in agreement with previous experimental observation of energy shift in ferroelectric/TMD structure~\cite{salazar_visualizing_2022}. Secondly, we further confirm that in this energy window, there is large range from -2 to -1.4 eV where the spin up accumulation becomes larger than the down one, as depicted in inset of Fig.~\ref{Fig4}b, in qualitative agreement with our experimental results. The photon energy of our pump laser ($\sim$1.55eV) corresponds as well to this energy window in the valence band. Interestingly, our results also indicate that the up/down ratio and thus the THz modulation depth could be increased up to 2 by acting on the Fermi level position. Finally, we further confirm that the SCC occurs at MoSe$_2$/PtSe$_2$ interface as the spin accumulation in pristine PtSe$_2$ is negligible (see Suppl. Info.~\ref{sec:S_acc_pristine}), in agreement with previous reports~\cite{Mudgal2023}.

This net spin accumulation results from the existence of Rashba states which are clearly visible when plotting the spin texture in that energy range (Fig.~\ref{Fig4}c). We thus interpret the observed THz emission change in up and down areas to arise from SCC mediated by IREE at the interface between MoSe$_2$ and PtSe$_2$, and modulated by ferroelectricity. This effect results from a positive spin accumulation and gives the same sign as in the case of Pt, as shown experimentally (Fig.~\ref{Fig3}c) and confirmed theoretically (see Suppl. Info.~\ref{sec:S_conv_sign}).

\section*{\label{sec:Conclusion} Conclusion.}
In conclusion, we have presented a large scale heterostructure made of FM/TMD junction on a bi-domain ferroelectric substrate. Using time-domain THz spectroscopy, we found that the THz emission could be modulated by the direction of the out-of-plane ferroelectric polarization. From DFT simulations, we found that the SCC is driven by IREE at the MoSe$_2$/PtSe$_2$ interface, and that the electric field direction allows to shift the semi-conducting bands, resulting in a sizeable change of THz emission. Our results show that the modulation depth could be further increased by varying the Fermi level position, which is reachable by gating the system and applying a voltage to the 2D layers. These findings open very interesting perspectives for the realization of efficient THz modulators based on field effect spin-orbit devices.

\begin{acknowledgments}
The authors acknowledge funding from European Union’s Horizon 2020 research and innovation program under grant agreement No 964735 (FET-OPEN EXTREME-IR). The French National Research Agency (ANR) is acknowledged for its support through the ANR-21-CE24-0011 TRAPIST, ANR-22-EXSP-0003 TOAST, ANR-22-EXSP-0009 SPINTHEORY and ESR/EQUIPEX+ ANR-21-ESRE-0025 2D-MAG projects. The authors also thanks F. Ibrahim, V. Libor and M. Chshiev from SPINTEC for their fruitful help into the calculation of electric field polarization in the 2D materials.
\\
The authors declare no competing interests.
\end{acknowledgments}

\bibliography{biblio_2D_LNO}

\appendix

\title{Supplementary Material of "Inverse Rashba Edelstein THz emission modulation induced by ferroelectricity in CoFeB/PtSe$_2$/MoSe$_2$//LiNbO$_3$ systems".}
\date{\today}

\author{S.~Massabeau}
\affiliation{Laboratoire Albert Fert, CNRS, Thales, Université Paris-Saclay, 91767 Palaiseau, France}

\author{O.~Paull}
\affiliation{Laboratoire Albert Fert, CNRS, Thales, Université Paris-Saclay, 91767 Palaiseau, France}

\author{A.~Pezo}
\affiliation{Laboratoire Albert Fert, CNRS, Thales, Université Paris-Saclay, 91767 Palaiseau, France}

\author{F.~Miljevic}
\affiliation{Laboratoire Albert Fert, CNRS, Thales, Université Paris-Saclay, 91767 Palaiseau, France}

\author{M.~Mičica}
\affiliation{Laboratoire de Physique de l’Ecole Normale Supérieure, ENS, Université PSL, CNRS, Sorbonne Université,
Université Paris Cité, F-75005 Paris, France}

\author{A. Grisard}
\affiliation{Thales Research \& Technology, 91767 Palaiseau, France}

\author{P. Morfin}
\affiliation{Laboratoire de Physique de l’Ecole Normale Supérieure, ENS, Université PSL, CNRS, Sorbonne Université,
Université Paris Cité, F-75005 Paris, France}

\author{R. Lebrun}
\affiliation{Thales Research \& Technology, 91767 Palaiseau, France}

\author{H.~Jaffrès}
\affiliation{Laboratoire Albert Fert, CNRS, Thales, Université Paris-Saclay, 91767 Palaiseau, France}

\author{S. Dhillon}
\affiliation{Laboratoire de Physique de l’Ecole Normale Supérieure, ENS, Université PSL, CNRS, Sorbonne Université,
Université Paris Cité, F-75005 Paris, France}

\author{J.-M.~George}
\affiliation{Laboratoire Albert Fert, CNRS, Thales, Université Paris-Saclay, 91767 Palaiseau, France}

\author{M.~Jamet}
\affiliation{Univ.~Grenoble Alpes, CEA, CNRS, Grenoble INP, IRIG-SPINTEC, F-38000 Grenoble, France}

\author{M.~Bibes}
\affiliation{Laboratoire Albert Fert, CNRS, Thales, Université Paris-Saclay, 91767 Palaiseau, France}

\maketitle

\begin{sloppypar}

\newpage

\section{\label{sec:S_surface}Surface characterization}
\begin{figure}[!ht]
  \centering
  \includegraphics[width=10cm]{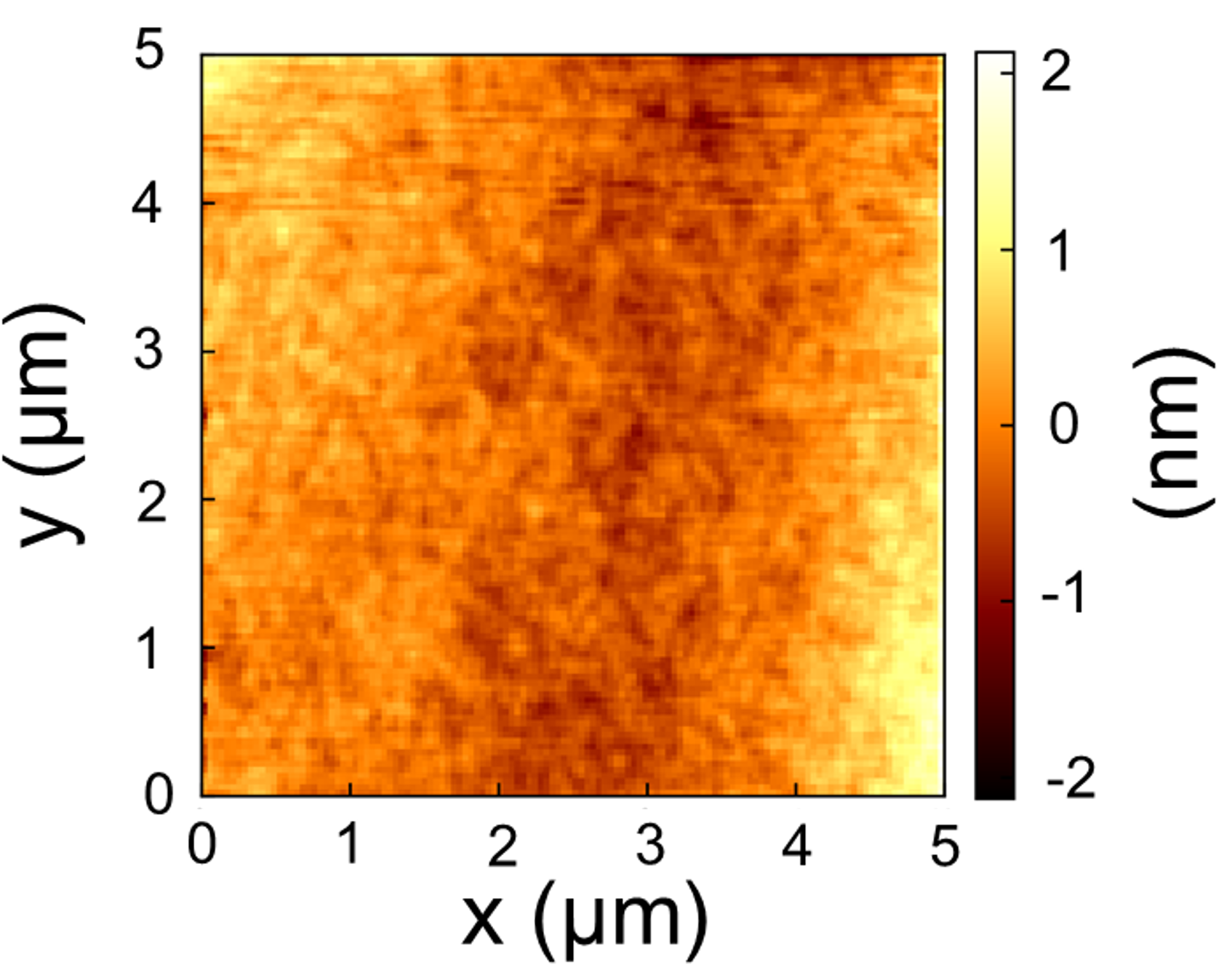}
  \caption{Atomic force microscopy of the surface of the LiNbO$_3$ substrate before
 transfer of the 2D layer.}
  \label{Fig_S_AFM}
\end{figure}

In order to ensure the MoSe$_2$/PtSe$_2$ interface quality after the transfer process, We first check on the LiNbO$_3$ surface quality before transfer. As depicted in Fig.~\ref{Fig_S_AFM}, we found a very low roughness with a RMS of 0.33 nm.

\section{\label{sec:S_spin_acc}Spin accumulation and spin texture calculations}

 DFT simulations were carried out using the Perdew-Burke-Ernzerhof (PBE) \cite{gga,pbe} exchange-correlation functional. We perform the geometry optimizations employing a localized atomic basis based on SIESTA~\cite{siesta_method} with an atom-centered double-$\zeta$ plus polarization (DZP) basis sets considering an energy cut-off for the real-space mesh of 400 Ry. Relativistic effects were introduced via the off-site approximation \cite{siesta_off-site_soc} using fully-relativistic norm-conserving pseudopotentials \cite{tm_pseudopotentials}. The system's ground state properties are obtained after performing a full self-consistent cycle converged on a $(15 \times 15 \times 1)$  $\vec{k}$-points sampling of the Brillouin zone, using a 25 {\AA} vacuum distance between periodic images to avoid spurious effects. The lattice constant is fixed to that of PtSe$_2$, corresponding to a $2 \times 2$ supercell of size 6.57 {\AA} having a $\sqrt{3} \times \sqrt{3}$ MoSe$_2$ supercell. We used the conjugate gradient algorithm to minimize the atomic forces below 0.01 eV/{\AA} with the inclusion of VdW corrections as implemented in the DFT code. 

The spin texture shown in Fig. 4c was calculated using:

\begin{equation*}
    \braket{\mathbf{S}}_n=\braket{\psi_n|\mathbf{\hat{S}}|\psi_n},
\end{equation*}

where $\ket{\psi_n}$ is an eigenstate of the Hamiltonian and the Overlap  matrix $\hat{S}$ has to be considered as the basis is non-orthogonal. The matrices $\hat{\mathbf{S}}$ are the usual spin operator matrices proportional to the Pauli matrices. Analogously, the spin accumulations was obtained by using linear response theory:

\begin{equation}
  \braket{\hat{S}}_y=   -e E_x \tau\int_{BZ} \partial_\epsilon f(\epsilon)d\epsilon \operatorname{Re}\left[ \braket{\psi_n|\hat{S}_y|\psi_n}\braket{\psi_n|v_x|\psi_n}\right]
\label{eq:accum}
\end{equation}

where $v_x$ is the velocity operator modified for treating a non-orthogonal basis as:

\begin{equation}
    \mathbf{v}= \frac{\partial
    \hat{\mathbf{H}}}{\partial \mathbf{k}}
    -\frac{\varepsilon_i \partial
    \hat{\mathbf{S}}}{\partial \mathbf{k}},
\end{equation}

which for our case was taken in the $\hat{x}$ parallel to bilayer structure representing the external perturbation and $\hat{S}_y$ is the operator corresponding to the spin degree of freedom. In order to achieve convergence in the transport calculations we have considered a $150\times 150$ mesh-grid for sampling the Brillouin Zone (BZ) where \ref{eq:accum} was integrated.

\section{\label{sec:S_band_structure}Band structure of the trilayer system}

\begin{figure}[!ht]
  \centering
  \includegraphics[width=16cm]{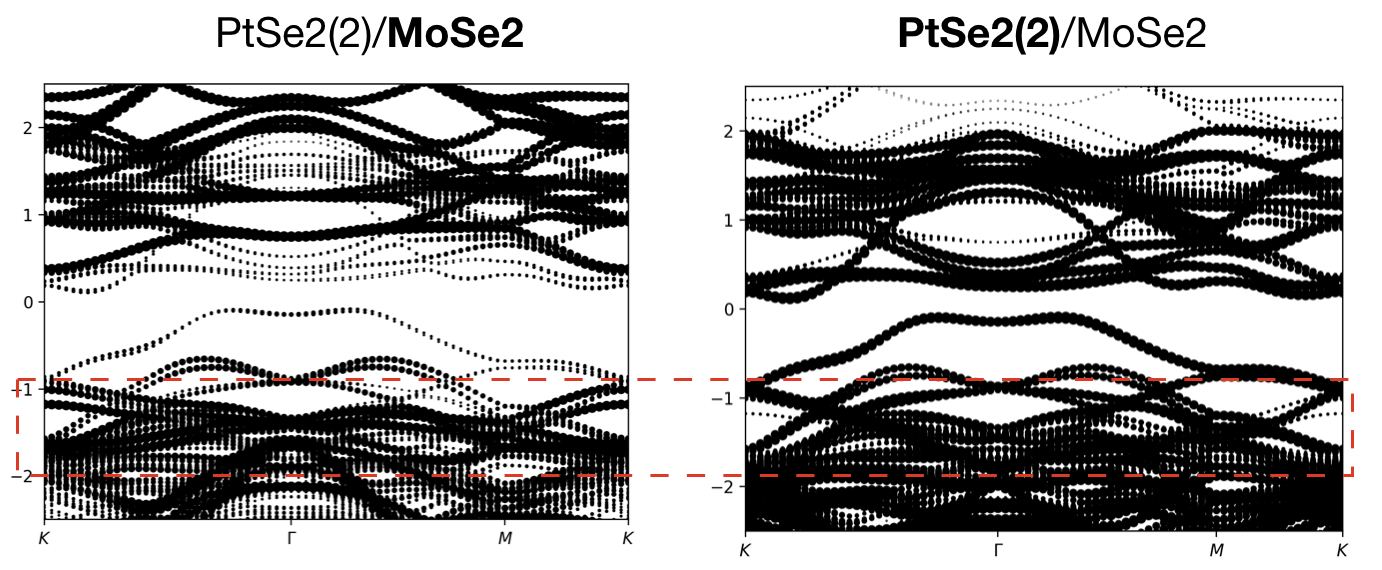}
  \caption{Band structure of the MoSe$_2$(1ML)/PtSe$_2$(2ML) trilayer, projected either on MoSe$_2$ (left) or PtSe$_2$ (right) layers. Dashed red rectangle highlights the energy window with an increased density of states in MoSe$_2$(1ML).}
  \label{FigS2}
\end{figure}

We present on Fig.~\ref{FigS2} the band structure of the MoSe$_2$(1ML)/PtSe$_2$(2ML) trilayer, projecter either on MoSe$_2$ (left) or PtSe$_2$ (right). From our DFT simulations, we see that by considering a PtSe$_2$ bilayer we get a reduced gap, in agreement with previous studies \cite{Kandemir2018}. This would allow spin-to-charge conversion near the Fermi level, provided the existence of PtSe$_2$ states. On the other hand, when MoSe$_2$ is considered, overlap between states coming from both systems occur only below -1 eV (dashed red rectangle) and for higher energies above the Fermi level. This characteristic is also highlighted by calculating the projected density of states, as shown in Fig. 4.a in the main text.

\section{\label{sec:S_acc_pristine}Spin accumulation in pristine PtSe2}

\begin{figure}[!ht]
  \centering
  \includegraphics[width=15cm]{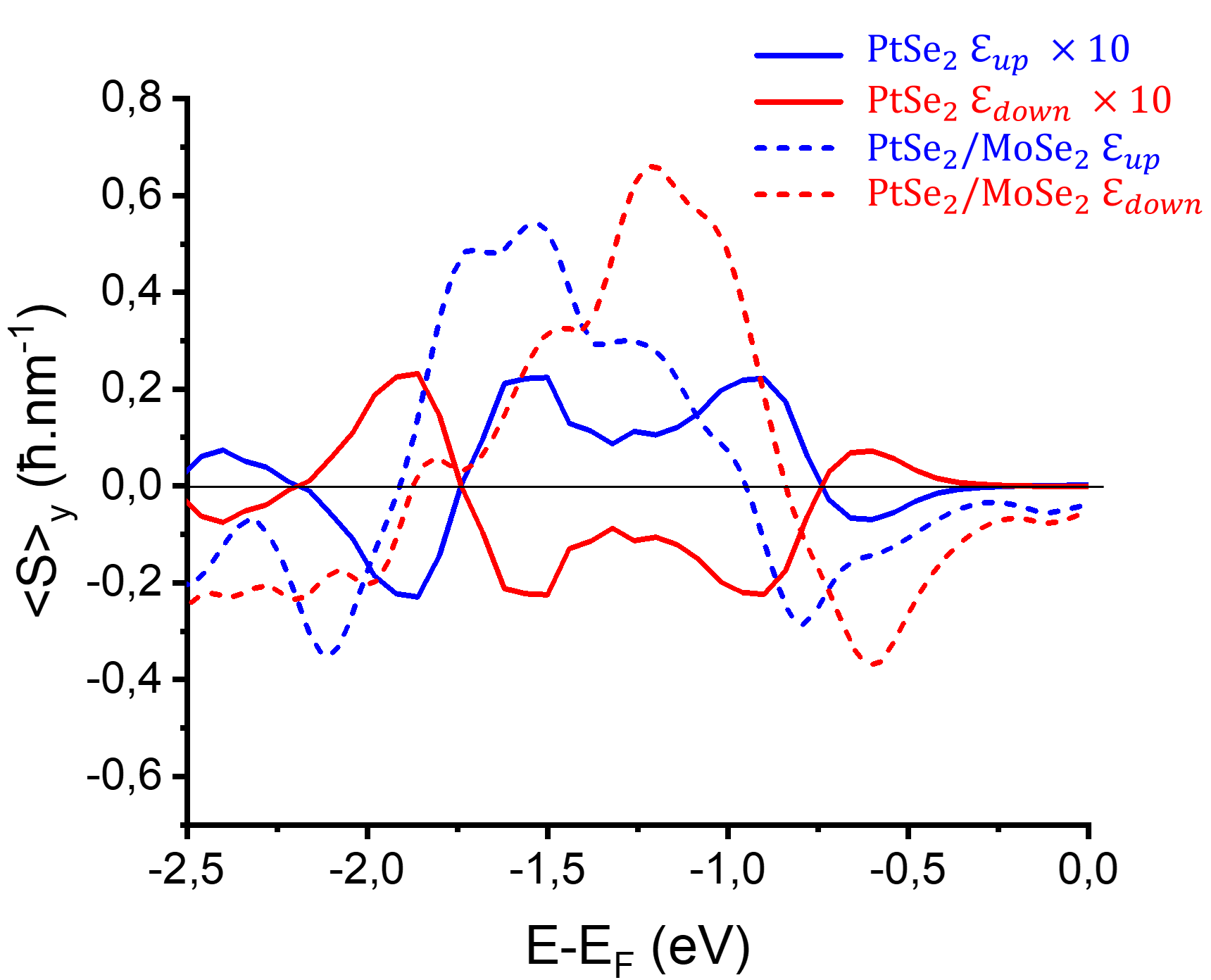}
  \caption{Spin accumulation as a function of the energy for pristine PtSe$_2$ (plain lines) and MoSe$_2$(1ML)/PtSe$_2$(2ML) trilayer (dashed lines), while applying an external electric field $\mathcal{E}$=0.25 V/{\AA} pointing up (blue) and down (red), in agreement with the experiment. The spin accumulation in the pristine case is about one order of magnitude lower than in the trilayer case.}
  \label{FigS_pristine}
\end{figure}

In Fig.~\ref{FigS_pristine}, we show the calculation of spin accumulation, related to the Rashba Edelstein tensor, for pure PtSe$_2$ when an electric field is applied, using the same electric field as that in the main text for the trilayer system MoSe$_2$(1ML)/PtSe$_2$(2ML). Although the inversion symmetry breaking on PtSe$_2$ leads to a non-vanishing value of the spin accumulation (plain lines), its value is at least one order of magnitude smaller than the effect when MoSe$_2$ is added (dashed lines), meaning that the main effect responsible for the SCC in the trilayer system comes from the hybridized orbitals between PtSe$_2$ and MoSe$_2$. Furthermore, we clearly see a reversal of the REE tensor when the electric field is reversed in the case of pure PtSe$_2$, since the Rashba effect captures the chirality related to the direction on which inversion symmetry is broken. This is contrast with the situation depicted in the MoSe$_2$(1ML)/PtSe$_2$(2ML) trilayer where there's no such reversal in the sign of the tensor. This result is in agreement with previous report on negligible Rashba effect in FM/PtSe$_2$ structures~\cite{Mudgal2023}, and further corroborates our assumption that SCC happens in the valence bands~\cite{abdukayumov_atomiclayer_2024} due to a strong orbital hybridization between PtSe$_2$ and MoSe$_2$.

\section{\label{sec:S_conv_sign}Conversion sign comparison with Pt and Co}

In order to compare the conversion sign obtained from spin accumulation modelling in MoSe$_2$(1ML)/PtSe$_2$(2ML) systems, we calculate the spin Hall conductivity of Pt and W. In the case of systems made of FM/Pt and FM/W, the SCC is driven by Inverse Spin Hall Effect, and the THz emission is directly proportionnal to the spin Hall conductivity, with a positive value for Pt (sets by convention) and an opposite one (negative) for W.

From our calculation, we recover Pt spin Hall conductivity to be positive, with W exhibiting an opposite phase, nearby the Fermi level. As we experimentally found SCC in our heterostructure to be the same sign as of Pt, we conclude that positive values for spin accumulation define the relevant energy window involved in the conversion process.

\begin{figure}[!ht]
  \centering
  \includegraphics[width=16cm]{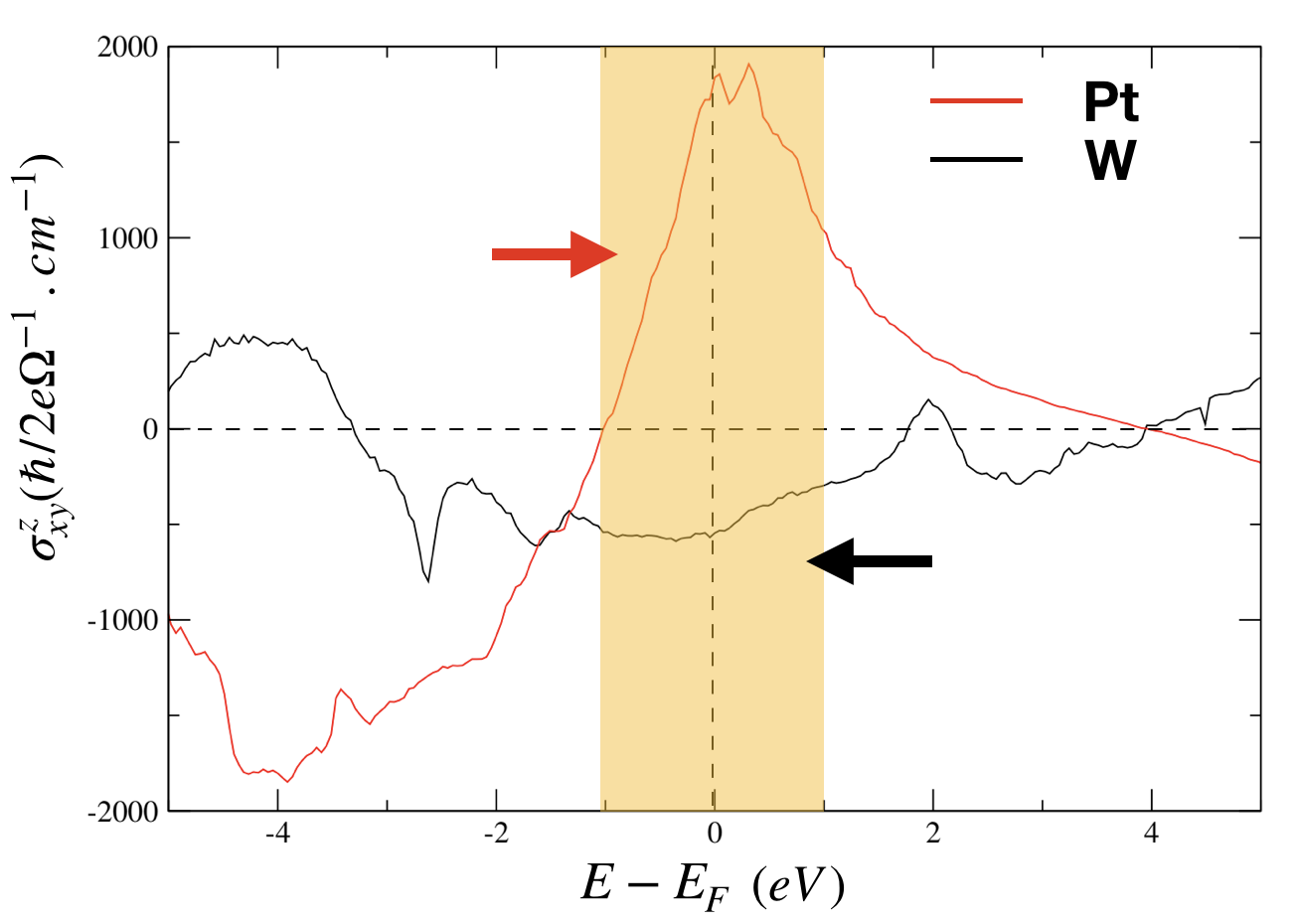}
  \caption{Spin Hall conductivity $\sigma_{x,y}^{z}$ as a function of the energy for Pt (red) and W (black), showing inverse signs nearby the Fermi level (orange area). Pt sign is found to be positive, in agreement with convention, which thus allows to conclude on the emission sign from FM/TMD heterostructure.}
  \label{FigS3}
\end{figure}

\end{sloppypar}

\end{document}